\documentclass[twocolumn,american,amssymb,amsmath,superscriptaddress,aps,pre,reprint]{revtex4-1}
\usepackage[latin9]{inputenc}
\usepackage{array}
\usepackage{amsmath}
\usepackage{graphicx}

\makeatletter

\providecommand{\tabularnewline}{\\}

\usepackage{dcolumn}
\usepackage{bm}
\usepackage{subfigure}

\makeatother

\usepackage{babel}
\begin{document}

\title{Mean field dynamics of a random neural network with noise}

\author{Vladimir Klinshov}

\email{vladimir.klinshov@ipfran.ru}

\affiliation{Institute of Applied Physics of the Russian Academy of Sciences,
46 Ulyanov Street, 603950 Nizhny Novgorod, Russia}

\author{Igor Franovi\'{c}}

\email{franovic@ipb.ac.rs }

\affiliation{Scientific Computing Laboratory, Institute of Physics Belgrade, University
of Belgrade, Pregrevica 118, 11080 Belgrade, Serbia}
\begin{abstract}
We consider a network of randomly coupled rate-based neurons influenced
by external and internal noise. We
derive a second-order stochastic mean-field model for the network dynamics and use it to analyze the stability
and bifurcations in the thermodynamic limit, as well as to study the fluctuations due to the finite-size effect. It is demonstrated
that the two types of noise have substantially different impact on
the network dynamics. While both sources of noise give rise to stochastic
fluctuations in case of the finite-size network, only the external
noise affects the stationary activity levels of the network in the
thermodynamic limit. We compare the theoretical predictions with the
direct simulation results and show that they agree for large enough
network sizes and for parameter domains sufficiently away from bifurcations. 
\end{abstract}
\maketitle

\section{Introduction}

The different stages of information processing in large neural systems
comprise multiple characteristic spatial and temporal scales. While
the \emph{in-vivo} recordings of single neurons indicate considerable
subthreshold fluctuations and highly variable spike trains \cite{SK93,STG01,DRP03},
the macroscopic measurements have revealed reliable and structured
activity in many cortical areas \cite{B06a}. Accounting for these
two results is an outstanding theoretical issue, which requires one
to develop analytically tractable models capable of capturing the
functional organization and integration of single unit dynamics at
different levels of complexity. This is typically resolved by invoking
the mean-field approach to describe the coarse-grained activity and
interactions of neural populations. Given the often used assumption
on population homogeneity, the approach is from biological view most
appropriate for intermediate-scale (mesoscopic) assemblies, such as
cortical columns \cite{M57,GZTK08}. The latter assemblies incorporate
on one hand a sufficiently large number of neurons for the averaging
effects to occur, but on the other hand, are small enough to support
the homogeneity assumption.

The mean field approach has so far been implemented to network structures
as well as spatially extended neural systems, with the pertaining
models classified as activity-based or voltage-based depending on
the type of the state variable \cite{CGPW14,GZTK08}. The seminal
works of Wilson and Cowan \cite{WC72,WC73}, as well as Amari \cite{A77},
employed the heuristic continuum limit, providing the description
of the temporal coarse-grained dynamics in neural fields. Though deterministic
in nature, such models recovered a number of highly relevant dynamical
regimes including multistability \cite{WC72,WC73,A77}, large-scale
oscillations \cite{B06a,RRWBGR01,JH97}, stationary pulses or bumps
\cite{A77,LC01,LTGE02}, traveling fronts and pulses \cite{PE01,FB05b,EB03},
spiral waves \cite{L05} as well as spatially localized oscillations
\cite{FB05a,OLC07}. Nevertheless, given the aim to reconcile observations
of highly variable local neuron activity and the substantially reliable
activity patterns at the macroscopic scale, the key point emerging
in recent research on mean-field models has been to account for the
higher-order statistics \cite{B09,B10,T12,TE11}. Conceptually, the
goal has become to demonstrate how the fluctuations and correlations
from the single unit level translate to and are manifested at the
assembly level.

In general, the physical background of variability of single units
may be related either to noise or the balanced recurrent excitatory
and inhibitory inputs \cite{RMBWP07,MBP06,RD14}. Our interest lies
with the former scenario. In neural systems, noise may derive from
a number of extrinsic and/or intrinsic sources \cite{FSW08,DW11,AANVSG07,HJBS06,JBS08}.
The external noise is mainly due to random inputs arriving from a
large number of afferent neurons (synaptic noise), whereas the internal
noise is primarily linked to random opening of a finite number of
ion channels (ion-channel noise). In the present paper, we consider
a network of randomly connected units, where the local dynamics follows
a rate-model and is affected both by the internal and the external
noise. Using the Gaussian closure hypothesis \cite{LGNS04,FTVB13,FTVB14},
we will derive the stochastic mean-field model characterizing the
macroscopic network activity in terms of the mean rate and the associated
variance.

The issue of how noise from the single unit level translates to noise
at the macroscopic scale is highly nontrivial. So far, the stochastic
mean-field models have been constructed either via the top-down or
the bottom-up approaches. In the top-down approach, the details of
the local neuron dynamics are neglected, which typically leads to
phenomenological stochastic neural field models. These are based either
on Langevin version of the deterministic equations, having introduced
some form of spatiotemporal white noise \cite{HLSG08,FTC09}, or on
treating the neural field equations as the thermodynamic limit of
the underlying master equation \cite{CGPW14,B10,BD09}. In the latter
case, extensions of the deterministic mean-field model have been obtained
by perturbation techniques, such as the system-size expansion \cite{B09,K92},
or via the field-theory methods, viz. the path integral formalism
\cite{BC07,BCC10}. The bottom-up construction of stochastic mean-field
models has primarily concerned networks of integrate-and-fire neurons
with two types of interaction topology, the global coupling scheme
\cite{GH93} or the sparse connectivity \cite{BH99,B00}. Within the
framework of population density method \cite{BH99,B00,AV93,NT00,LT07},
such networks have been shown to display the asynchronous state despite
the fact that the local firing conforms to Poissonian process. Under
such conditions, the collective dynamics has been described by an
effective mean-field rate equation with a characteristic gain function.
Nevertheless, the point that the asynchronous state is stable only
in the thermodynamic limit has indicated that the finite-size effects
\cite{SC07,GS94,MV02} may yield qualitatively novel phenomena and
contribute as additional source of intrinsic noise at the network
level.

Apart from considering the networks of spiking neurons, the bottom-up
approaches to stochastic mean-field models have pursued the second
line of research featuring local rate dynamics \cite{H07a,H07b,H09}.
This is consistent with the long standing debate on the precise temporal
codes vs. rate codes as the main principles of information encoding
in neural systems \cite{UR99,CZ00}. The importance of rate code has
been confirmed for a number of motor and sensory areas \cite{AMP04,CMMN99},
whereby the potential advantage of the population rate code may lie
in the lesser vulnerability to noise. For the class of models built
on the rate-based neurons, Hasegawa has introduced the augmented moment
approach \cite{H07a,H07b,H03} to analyze the mean-field dynamics
of globally coupled finite-size populations where the units are subjected
to additive and multiplicative noise. While we also consider the rate-based
neurons, our model is distinct in that it accounts for the effects
arising from the random network topology. Also, the issue of how the
effects of noise acting on single units are manifested at the assembly
level is addressed in a more elaborate fashion, accounting for the
origin of multiplicative noise in the mean-field dynamics.

The paper is organized as follows. In Sec. \ref{sec:Derivation},
we introduce the rate model of local activity and apply the Gaussian
closure hypothesis to derive the stochastic mean-field equations for
the finite-size population of randomly connected units. Section \ref{sec:Thermo}
concerns the stability analysis of the introduced mean-field model
in the thermodynamic limit, where noise intensities act as additional
system parameters. Apart from demonstrating the emergence of macroscopic
bistable behavior, it is also shown how temporary changing of the
level of noise may be used to control the network state in a hysteresis-like
scenario. In Sec. \ref{sec:Finite} we discuss the finite-size effects
and determine the magnitude of fluctuations around the stationary
states from the thermodynamic limit. Section \ref{sec:Summary} provides
a brief summary of the results obtained.

\section{Derivation of the mean-field model\label{sec:Derivation}}

We consider a network of $N$ excitatory neurons. The local activity
is described in terms of firing rates $r_{i},i\in[1,N]$, whose dynamics
is given by: 
\begin{equation}
\frac{dr_{i}}{dt}=-\lambda r_{i}(t)+\mathcal{H}\left(\kappa u_{i}(t)+I+\sqrt{2B}\zeta_{i}(t)\right)+\sqrt{2D}\xi_{i}(t).\label{eq:1}
\end{equation}
In the last equation, $\lambda$ denotes the relaxation characterizing
the inertness of units, $\mathcal{H}(u)$ is the gain function and
$\kappa=c/N$ stands for the coupling coefficient, and $I$ is the
external current which is taken to be constant. The above form of
rate model is considered paradigmatic \cite{H07a,AMP04}, and a substantial
amount of theoretical work has been carried out to analytically obtain
the particular transfer functions for a range of spiking neuron models
\cite{B00,FB02,SHS03}.

\begin{figure}[t]
 \includegraphics[width=7cm]{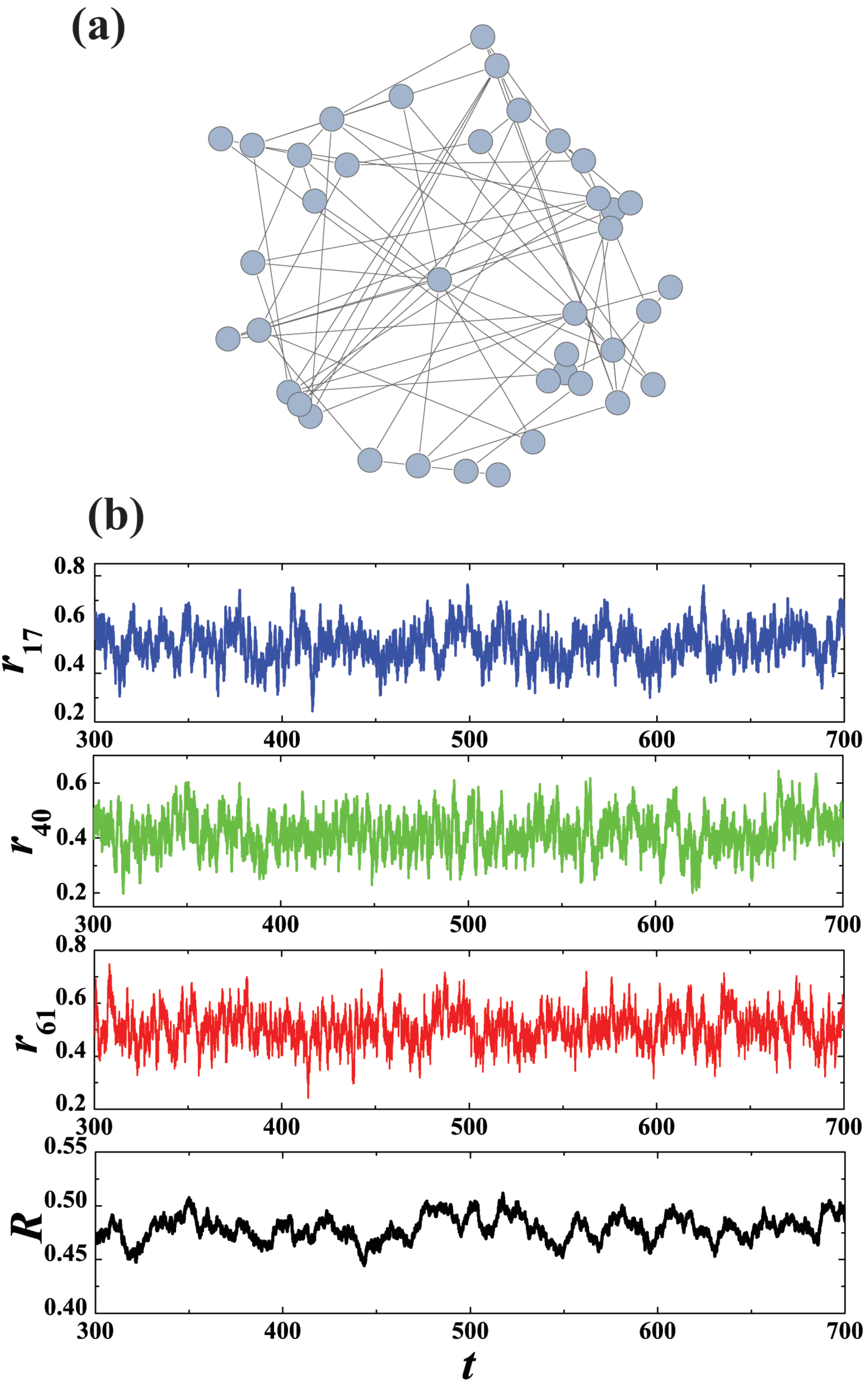} \protect\caption{(Color online) Illustration of the network topology and the typical
network activity. (a) shows a sample configuration of an Erd\"os-R\'enyi
network for $N=40$ and $p=0.1$. (b) Typical time series $r_{i}(t)$
of three arbitrary units (first three rows) are compared to the time
series of the mean rate $R(t)$ (bottom row). Note that the fluctuations
of the mean rate are much smaller than those for the local variables.
The system parameters are $N=300$, $p=0.2$, $c=3$, $B=0.002$,
$D=0.0005$ , $I=0.21$.\label{fig:net}}
\end{figure}

Each unit is influenced by the external (synaptic) white noise $\xi_{i}(t)$
and the internal (ion-channel) white noise $\zeta_{i}(t)$, whose
respective intensities are $B$ and $D$. The external and internal
noise sources are assumed to be independent, whereas the random perturbations
acting on different units are uncorrelated. The input $u_{i}$ which
the neuron $i$ receives from the rest of the network is specified
by 
\begin{equation}
u_{i}(t)=\sum_{j}a_{ij}r_{j}(t),\label{eq:2}
\end{equation}
where $a_{ij}\in\{0,1\}$ denote the elements of the adjacency matrix.
Throughout the paper, it is assumed that the interaction topology
is random, conforming to the Erd\"os-R\'enyi type of network, see
Fig. \ref{fig:net}(a).  

In the remaining part of this section, we derive the mean-field model
for the collective dynamics of the network given by the system \eqref{eq:1}-\eqref{eq:2}.
Our approach is essentially based on the well-known Quasi-independence
and Gaussian approximations \cite{FTVB14}, and leads to the second-order
mean-field model of the macroscopic dynamics. In other words, we use
the moment approach with the Gaussian closure hypothesis. The collective
behavior is then described in terms of the mean (assembly-averaged)
rate and the associated variance 
\begin{align}
R(t) & =\langle r_{i}\rangle\equiv\frac{1}{N}\sum_{i}r_{i}\nonumber \\
S(t) & =\langle(r_{i}(t)-R(t))^{2}\rangle=\langle r_{i}(t)^{2}\rangle-R(t)^{2}.\label{eq:2.5}
\end{align}
One naturally expects that the fluctuations of the mean-rate will
be comparably smaller than the fluctuations for the local variables,
see the sample series in Fig. \ref{fig:net}(b). This point will be
confirmed during the derivation of the mean-field equations. In order
to make the reading easier, a summary of the most relevant notation
used throughout the paper is provided in Table \ref{tab:notation}.

\begin{table}[t]
\protect\caption{\label{tab:notation}Summary of the introduced notation.}

\begin{ruledtabular}
\begin{tabular}{lp{2in}}
$\lambda$  & Relaxation time of units\tabularnewline
$c$  & Coupling strength\tabularnewline
$\kappa\equiv c/N$  & Normalized coupling strength\tabularnewline
$I$  & External current\tabularnewline
$D$  & Intensity of internal noise\tabularnewline
$B$  & Intensity of external noise\tabularnewline
$p$  & Connection probability \tabularnewline
$\alpha\equiv cp$  & Connectivity parameter\tabularnewline
$n$  & Mean number of connections per unit\tabularnewline
$R$  & Mean (assembly-averaged) rate\tabularnewline
$S$  & Rate variance\tabularnewline
\end{tabular}\end{ruledtabular}

\end{table}

Before proceeding to the analytical part, let us explicitly state
the approximations relevant for the derivation of the mean-field model.
The first one concerns the requirement that the random variables $r_{i}(t)$
at any moment $t$ and for sufficiently large $N$ satisfy $\langle r_{i}\rangle\approx[r_{i}(t)]$,
where $[\cdot]$ denotes the expectation over the different stochastic
realizations. The mathematical background of this approximation lies
in the strong law of large numbers, which states that the sample average
$Y_{N}=N^{-1}\sum_{i=1}^{N}y_{i}$ of $N$ independently and identically
distributed random variables $y_{i}$ will almost surely converge
to the expectation $[y_{i}]$ for $N\rightarrow\infty$. The form
of convergence for large, but finite $N$ is specified by the central
limit theorem. In physical terms, the outputs of neurons $r_{i}$
can be considered unbiased if the distribution of the number of incoming
connections (connectivity degrees) over the population is sufficiently
narrow.

The second approximation is in a sense implicit for the validity of
the first one, and consists in the requirement that the correlation
between the outputs of neurons is negligible: $[r_{i}(t)r_{j}(t)]=[r_{i}(t)][r_{j}(t)].$
This is reasonably satisfied when the units share a small fraction
of common input from the network \cite{BH99,VS98}. Recall that we
consider random Erd\"os-R\'enyi networks where the probability of
connection between two neurons equals constant value $p$. For such
networks, the fraction of the shared input for two neurons is $p$,
while the coefficient of variation for the number of incoming connections
equals $\sqrt{(1-p)/pN}$. Both values are small for $N\gg pN\gg1$.
Thus, in large sparsely connected random networks the approximations
for the mean-field approach should be fulfilled. This provided, one
can represent the output of each neuron as 
\begin{equation}
r_{i}=R+\sqrt{S}\rho_{i},\label{eq:2.6}
\end{equation}
where $\rho_{i}$ are uncorrelated variables with zero mean and unit
intensity, cf. \cite{T98,B06}.

Proceeding to the derivation of the mean-field model, let us for simplicity
first introduce the notation $x_{i}=\kappa u_{i}+I+\sqrt{2B}\zeta_{i}$
for the total input to the $i$-th neuron. Using \eqref{eq:2.6},
the latter can be written as 
\begin{equation}
x_{i}=X+k\nu_{i}R+\kappa\sqrt{S}\sum_{j}a_{ij}\rho_{j}+\sqrt{2B}\zeta_{i},\label{eq:xi}
\end{equation}
where $n_{i}=\sum_{j}a_{ij}$ denotes the number of incoming connections
to the $i$-th neuron, $n=\langle n_{i}\rangle=pN$ is the mean number
of connections, $\nu_{i}=n_{i}-n$ and $X=\kappa nR+I$. The deviations
$\nu_{i}$ are of the order of $\sqrt{pN}$, and the independence
of the variables $\rho_{i}$ implies $\sum_{j}a_{ij}\rho_{j}\sim\sqrt{n}$.
Therefore, the second and the third term in the righthand side of
(\ref{eq:xi}) are of the order of $1/\sqrt{N}$, i.e. are small.
If the external noise $B$ is weak as well, the function $\mathcal{H}(x_{i})$
can be expanded into the Taylor series around $X$: 
\begin{align}
\mathcal{H}(x_{i}) & \approx\mathcal{H}(X)+\mathcal{H}^{\prime}(X)(x_{i}-X)+\frac{1}{2}\mathcal{H}^{\prime\prime}(X)(x_{i}-X)^{2}=\nonumber \\
= & H_{0}+\kappa H_{1}\nu_{i}R+H_{2}\left(\kappa^{2}\nu_{i}^{2}R^{2}+\kappa^{2}Sn_{i}+2B\right)+\nonumber \\
+ & \left(H_{1}+2\kappa H_{2}\nu_{i}R\right)\left(\kappa\sqrt{S}\sum_{j}a_{ij}\rho_{j}+\sqrt{2B}\zeta_{i}\right).\label{eq:Hi}
\end{align}
In the last expression, we have introduced the notation $H_{0}=\mathcal{H}(X)$,
$H_{1}=\mathcal{H^{\prime}}(X)$ and $H_{2}=\frac{1}{2}\mathcal{H^{\prime\prime}}(X)$.
Note that the products of noisy terms are replaced by the respective
means: $\rho_{i}\rho_{j}=\delta_{ij},$ $\zeta_{i}\zeta_{j}=\delta_{ij},$
$\rho_{i}\zeta_{j}=0$.

Inserting \eqref{eq:Hi} into \eqref{eq:1}, one arrives at the equation
for the local rates 
\begin{equation}
\frac{dr_{i}}{dt}=-\lambda r_{i}+h_{i}+\gamma_{i}\sum_{j}a_{ij}\rho_{j}+\beta_{i}\zeta_{i}+\sqrt{2D}\xi_{i},\label{eq:ri}
\end{equation}
where $h_{i}=H_{0}+\kappa H_{1}R\nu_{i}+H_{2}\left(\kappa{}^{2}\nu_{i}^{2}R^{2}+\kappa^{2}Sn_{i}+2B\right)$,
$\gamma_{i}=\left(H_{1}+2\kappa H_{2}\nu_{i}R\right)\kappa\sqrt{S}$
and $\beta_{i}=\left(H_{1}+2\kappa H\nu_{i}R\right)\sqrt{2B}$. Taking
the population average of the equation for microscopic dynamics \eqref{eq:ri},
we obtain the following for mean (macroscopic) rate $R$:

\begin{eqnarray}
\frac{dR}{dt} & = & -\lambda R+H(R)+\frac{1}{N}\sum_{i,j}\gamma_{i}a_{ij}\rho_{j}+\nonumber \\
 &  & +\frac{1}{N}\sum_{i}\beta_{i}\zeta_{i}+\frac{\sqrt{2D}}{N}\sum_{i}\xi_{i}.\label{eq:R1}
\end{eqnarray}
where $H(R)=\langle h{}_{i}\rangle=H_{0}+H_{2}\left(\kappa^{2}M_{2}R^{2}+\kappa^{2}Sn+2B\right)$,
and $M_{2}=\langle\nu_{i}^{2}\rangle=p(1-p)N$ is the second central
moment of the connectivity degree distribution.

Note that Eq. \eqref{eq:R1} effectively includes three noisy terms.
Apart from the external and the internal noise, there is also the
``network noise'' due to variability in connectivity degrees. To
estimate the network noise, let us first rewrite the corresponding
term as $\frac{1}{N}\sum_{i,j}\gamma_{i}a_{ij}\rho_{j}=\sum_{j}\rho_{j}\frac{1}{N}\sum_{i}\gamma_{i}a_{ij}=\sum_{j}\rho_{j}\langle\gamma_{i}a_{ij}\rangle.$
Since $\gamma_{i}$ and $a_{ij}$ are not correlated, $\langle\gamma_{i}a_{ij}\rangle\approx p\langle\gamma_{i}\rangle$
holds. Taking this into account, the sum of noisy terms in (\ref{eq:R1})
can be rewritten as $\xi_{R}=\frac{1}{N}\sum_{i}\left(cpH_{1}\sqrt{S}\rho_{i}+\beta_{i}\zeta_{i}+\sqrt{2D}\xi_{i}\right)$,
which is equivalent to white noise with the intensity $2\Psi/N$,
where 
\begin{equation}
2\Psi=H_{1}^{2}c^{2}p^{2}S+2BH_{1}^{2}+2D.\label{eq:DR}
\end{equation}
In the last expression, the terms of the order of $1/N^{2}$ have
been neglected.

Now let us derive the equation for the variance $S$. Taking the appropriate
It\={o} derivatives, one obtains 
\[
\frac{dS}{dt}=\langle2r_{i}\frac{dr_{i}}{dt}+\gamma_{i}^{2}n_{i}+\beta_{i}^{2}+2D\rangle-2R\frac{dR}{dt}-2\Psi/N.
\]
It can readily be shown that the noisy terms completely cancel each
other. Using the assumption that the outputs $\rho_{i}$ are not correlated
to the connectivity $\nu_{i}$, one arrives at the following equation
for the variance 
\begin{align}
\frac{dS}{dt} & =2D+2BH_{1}^{2}-2\lambda S+\frac{1}{N}8p(1-p)BH_{2}^{2}c^{2}R^{2}\nonumber \\
 & -\frac{1}{N}p(1-p)H_{1}^{2}c^{2}S.\label{eq:S1}
\end{align}

Taking into account \eqref{eq:R1}, \eqref{eq:DR} and \eqref{eq:S1},
the stochastic mean-field model for the finite-size random network
of rate-based neurons reads: 
\begin{align}
\frac{dR}{dt} & =-\lambda R+H_{0}+2H_{2}B+\frac{c^{2}H_{2}}{N}\left(p(1-p)R^{2}+pS\right)\nonumber \\
 & +\sqrt{\frac{1}{N}\left(H_{1}^{2}c^{2}p^{2}S+2BH_{1}^{2}+2D\right)}\eta,\label{eq:R}\\
\frac{dS}{dt} & =2D+2BH_{1}^{2}+\frac{1}{N}8p(1-p)BH_{2}^{2}c^{2}R^{2}\nonumber \\
 & -\left(2\lambda+\frac{1}{N}p(1-p)H_{1}^{2}c^{2}\right)S.\label{eq:S}
\end{align}

Before proceeding with the stability and bifurcation analysis, a brief
remark is required regarding the numerical treatment of system \eqref{eq:1},
and the ensuing dynamics for the assembly average. In particular,
the transfer function involves an argument with the stochastic term
corresponding to external noise, which cannot be resolved unless some
approximation is introduced. During the derivation of the mean-field
model, we have expanded the transfer function $H(x_{i})$ to Taylor
series up to second order around the assembly-averaged input $X$,
having verified that each of the terms contributing the deviation
of the input $x_{i}$, received by an arbitrary unit $i$, from $X$
is small. The expansion up to second order may effectively be interpreted
as Gaussian approximation for the distribution of $H(x_{i})$ over
the assembly. When numerically integrating the system, one cannot
hold that such an approximation holds \textit{a priori}. It has to
be explicitly verified that the distribution of $H(x_{i})$ is indeed
Gaussian for the considered range of neuronal and network parameters.
To this end, before running the simulations, we have calculated the
distributions of the function $H(x+\sqrt{2B}\zeta)$ for various $x$
and evinced that their skewness and excess kurtosis are small, consistent
with the Gaussian requirement. This allowed us to replace the term
$H(x+\sqrt{2B}\zeta)$ by a Gaussian process with the same mean and
variance.

\begin{figure}[t]
\centering \includegraphics[width=8cm]{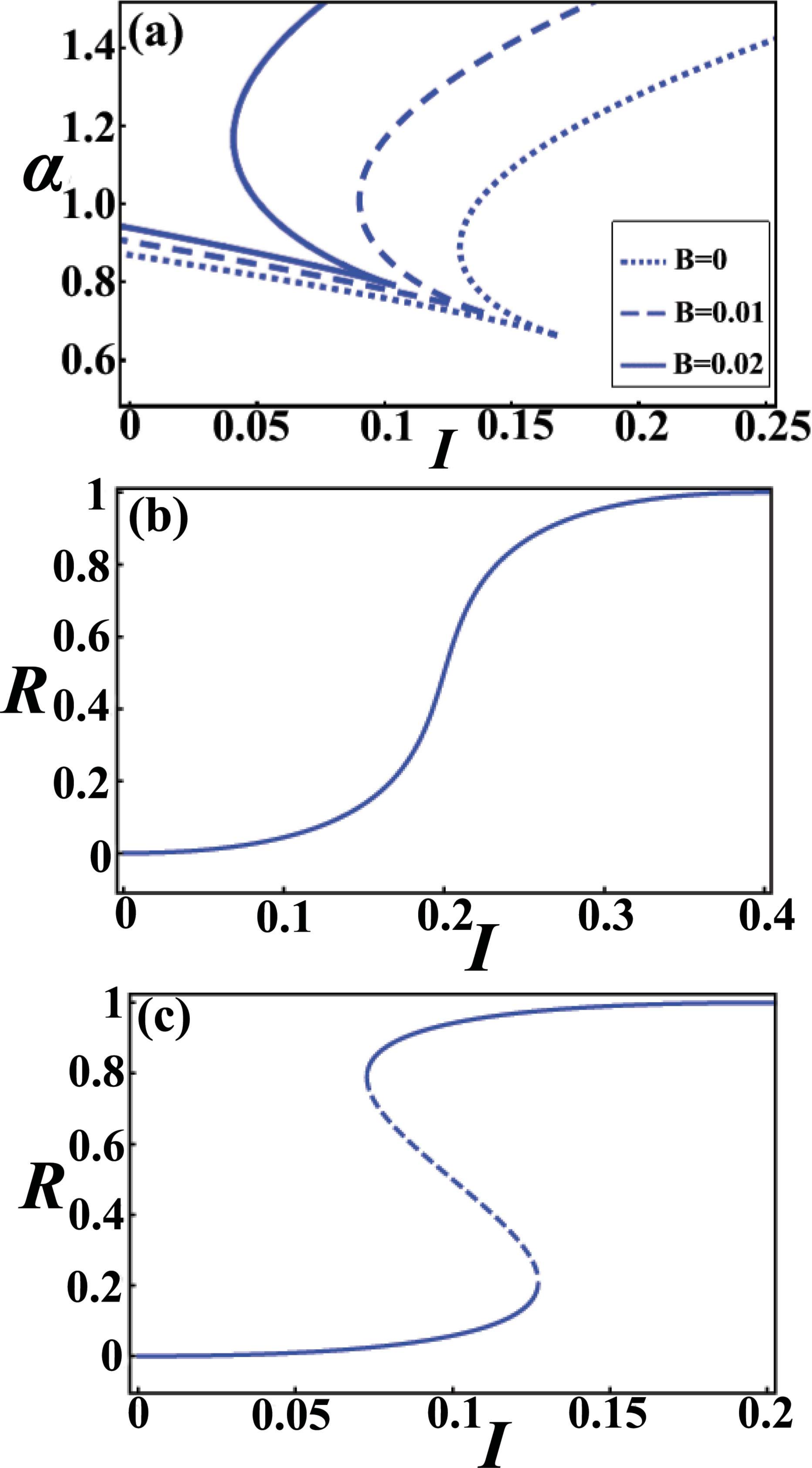}
\protect\protect\caption{(Color online) (a)Two-parameter bifurcation diagram of the network
in the thermodynamic limit. The lines show saddle-node bifurcations
in the $I-\alpha$ plane for three different values of $B$. (b) One-parameter
bifurcation diagram showing the dependence of the mean rate $R$ against
the bias current $I$ for $B=0$, $\alpha=0.6$. (c) The analogous
bifurcation diagram as in (b) is displayed for $\alpha=0.8$. The
solid lines indicate the stable branches, whereas the dashed line
refers to the unstable branch. \label{fig:thermo_bd}}
\end{figure}

\begin{figure*}[t]
\centering \includegraphics[scale=0.15]{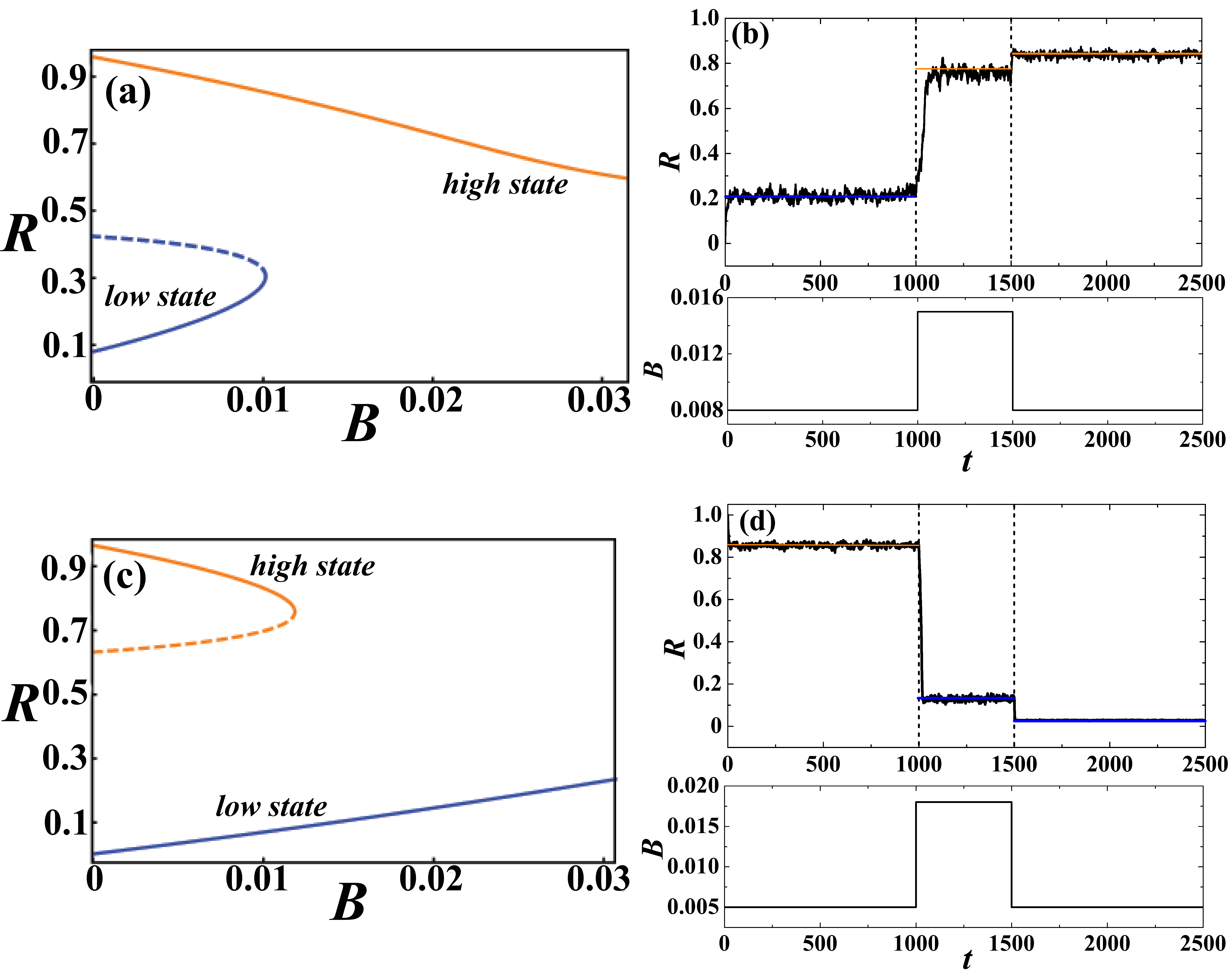} \protect\protect\caption{(Color online) (a) One-parameter bifurcation diagram illustrating
the dependence $R(B)$ for $\alpha=0.8$, $I=0.11$. The solid lines
indicate the stable branches of steady state solutions, whereas the
dashed line stands for the unstable branch. (b) The dynamics of the
network under temporary increase of external noise: $B=0.004$ for
$t<1000$ and $t>1500$, and $B=0.015$ for $t\in[1500;2000]$. The
black thick line refers to numerical results, whereas the thin solid
lines indicate the theoretically obtained stable activity levels for
the corresponding time intervals. (c) and (d) illustrate the switching
scenario from the high state to the low state by the temporary increase
of $B$. The network parameters are $\alpha=0.9$, $I=0.02$, whereas
$B$ values are $B=0.005$ for $t<1000$ and $t>1500$ and $B=0.018$
for $t\in[1500;2000]$. The presentation style is analogous to that
from (a) and (b). The network size in simulations is $N=300$.\label{fig:noise}}
\end{figure*}

\section{Analysis of stability and bifurcations in the thermodynamic limit\label{sec:Thermo}}

In this section, we analyze the stability and bifurcations of the
mean-field model (\ref{eq:R}-\ref{eq:S}) in the thermodynamic limit
$N\to\infty$. Under such conditions, the stochastic term in (\ref{eq:R})
can be neglected, so that the network dynamics effectively becomes
deterministic. The influence of noise is reduced to respective noise
intensities $B$ and $D$, which may be regarded as additional system
parameters. For simplicity, let us further set $\lambda=1$ and consider
the activation function $\mathcal{H}(x)$ of the form 
\begin{equation}
\mathcal{H}(x)=\begin{cases}
0, & x\leq0,\\
3x^{2}-2x^{3}, & 0<x<1,\\
1, & x\geq1.
\end{cases}\label{eq:H}
\end{equation}
Consistent with the notation introduced above, cf. \eqref{eq:Hi},
one has $H_{0}=3X^{2}-2X^{3}$, while the first- and second-order
derivatives are $H_{1}=6X-6X^{2}$, $H_{2}=6-12X$ for $0<X<1$.

The dynamics of variance $S$ in the thermodynamic limit is governed
by the equation 
\begin{equation}
\frac{dS}{dt}=2D+2BH_{1}^{2}-2S.\label{eq:S_thermo}
\end{equation}
Following relaxation, the variance reaches the stationary value 
\begin{equation}
S_{0}=D+BH_{1}^{2}.\label{eq:S0}
\end{equation}
Further note that the the dynamics of the mean rate $R$, given by
\eqref{eq:R}, becomes 
\begin{eqnarray}
\frac{dR}{dt} & = & -R+H_{0}+2H_{2}B\label{eq:R_thermo}
\end{eqnarray}
which is independent on the variance $S$. Taking into account that
$X=\alpha R+I$, where $\alpha=cp$, one can rewrite \eqref{eq:R_thermo}
as

\begin{eqnarray}
\frac{dX}{dt}=F(X) & = & -2\alpha X^{3}+3\alpha X^{2}-\left(12\alpha B+1\right)X+\nonumber \\
 &  & +6\alpha B+I.\label{eq:X_thermo}
\end{eqnarray}

The analysis of \eqref{eq:X_thermo} indicates that it always exhibits
at least one stable stationary state. For the parameter values given
by 
\begin{equation}
\alpha=\alpha_{0}=\frac{2}{3(1-8B)},\quad I=I_{0}=\frac{1-\alpha_{0}}{2},\label{eq:pitchfork}
\end{equation}


\eqref{eq:X_thermo} undergoes pitchfork bifurcation where two stable
steady states are created separated by an unstable one. The stable
states correspond to two distinct values of the mean firing rate which
we further refer to as the ``low'' and the ``high'' state. For
strong enough coupling $\alpha>\alpha_{0}$, the high (low) state
emerges via the saddle-node bifurcation, which occurs at the parameter
value 
\begin{equation}
I=\frac{1-\alpha}{2}\mp\frac{2}{3\sqrt{3}}\left(\frac{\alpha}{\alpha_{0}}-1\right)^{3/2},\label{eq:I}
\end{equation}
where the minus sign corresponds to the high, and plus to the low
state. For $I$ between these two values, the high and the low states
coexist, such that the network is in bistable regime. The two-dimensional
bifurcation diagram in Fig.\ref{fig:thermo_bd}(a) shows the curves
\eqref{eq:I} for different values of $B$. One can see that the two
curves form a ``tongue'' inside which the network is bistable. Figures
\ref{fig:thermo_bd}(b) and (c) display the one-dimensional bifurcation
diagrams for parameter values outside and within the bistability tongue,
respectively.

We note the interesting role played by the intensity of external noise
$B$. It is found to influence the position of the bistability region,
shifting it ``upwards'' toward the domain of stronger couplings.
This observation instigated an idea of the potential network control
mechanism via the noise intensity. In order to illustrate this mechanism,
we have analyzed in more detail how the network dynamics depends on
$B$. To this end, one can solve the equation $F(X)=0$ with respect
to $B$ and obtain the following expression 
\begin{equation}
B=\frac{1}{24}\left(\left(3-\frac{2}{\alpha}\right)-4\left(X-\frac{1}{2}\right)^{2}+\frac{\frac{2I-1}{\alpha}+1}{X-\frac{1}{2}}\right).
\end{equation}

For $I>\frac{1}{2}(1-\alpha)$, the corresponding one-dimensional
bifurcation diagram is provided in Fig. \ref{fig:noise}(a). The dependence
$X(B)$ is single-valued for $B>B_{0}$ , where $B_{0}=\frac{1}{24}\left(\left(3-\frac{2}{\alpha}\right)-3\left(\frac{2I-1}{\alpha}+1\right)^{2/3}\right)$.
For such $B$, only the high state of the network exists. For $B=B_{0}$,
the saddle-node bifurcation takes place, whereby the low state is
born. The latter state is found for $B<B_{0}$. Since only positive
values of $B$ are physically meaningful, the low state branch exist
only for $B_{0}>0$, which is equivalent to the condition 
\begin{equation}
\alpha>\frac{2}{3}\mbox{ and }\frac{1}{2}(1-\alpha)<I<\frac{1}{2}(1-\alpha)+\frac{\left(\alpha-2/3\right)^{3/2}}{\alpha^{1/2}}.\label{eq:cond}
\end{equation}
For $\alpha$ and $I$ satisfying \eqref{eq:cond}, the network is
bistable for $B<B_{0}$ and exhibits only the high state for $B>B_{0}$.
Therefore, a pulse-like increase of $B$ may switch the network from
the low to the high state via the hysteresis scenario.

This effect is illustrated in Fig. \ref{fig:noise}(b), which shows
the network dynamics before, during and after the pulse-like change
of the external noise $B$. Prior to strengthening, the external noise
level is $B=10^{-3}$, such that the network is bistable and is settled
in the low lying state. As soon as the external noise intensity is
temporarily increased to $B=5\times10^{-3}$, the low state vanishes,
and the network switches to the high state. When $B$ regains the
initial value, \eqref{eq:X_thermo} admits bistable regime again,
but the network remains in the high state. Thus, the temporary increase
of $B$ has caused the network to switch from the low state to the
high state.

Note that for $I<\frac{1}{2}(1-\alpha)$, the inverse scenario is
possible, where the network can switch from the high state to the
low state by a pulse-like increase of the external noise. An example
for such a scenario is illustrated in Fig. \ref{fig:noise}(c) and
Fig. \ref{fig:noise}(d).

Interestingly enough, the external noise has effect not only on the
stationary states of the network, but is found to influence its transient
dynamics as well. The transient dynamics is important when the external
input changes and the network has to track this change and adapt to
its rate accordingly. In this scenario, short response time of a network
is naturally considered as advantageous \cite{VS96}. Our analysis
shows that under certain conditions,introduction of the external noise
may sufficiently reduce the response time. To understand this, let
us consider the situation when the input $I$ switches from some value
$I_{1}$ to the new value $I_{2}$. For simplicity, we assume that
the other parameters are set so that the network is always monostable.
Then, the network will evolve from the previous stationary state $X_{1}$
to the new one $X_{2}$. According to (\ref{eq:X_thermo}), the rate
$\Gamma$ of the system convergence to $X_{2}$ is determined by the
absolute value of the derivative $F^{\prime}(X_{2})$:

\begin{equation}
\Gamma=-F^{\prime}(X_{2})=-6\alpha X_{2}^{2}+6\alpha X_{2}+12\alpha B+1.\label{eq:rate_thermo}
\end{equation}

\begin{figure}[t]
\centering \includegraphics[width=7cm]{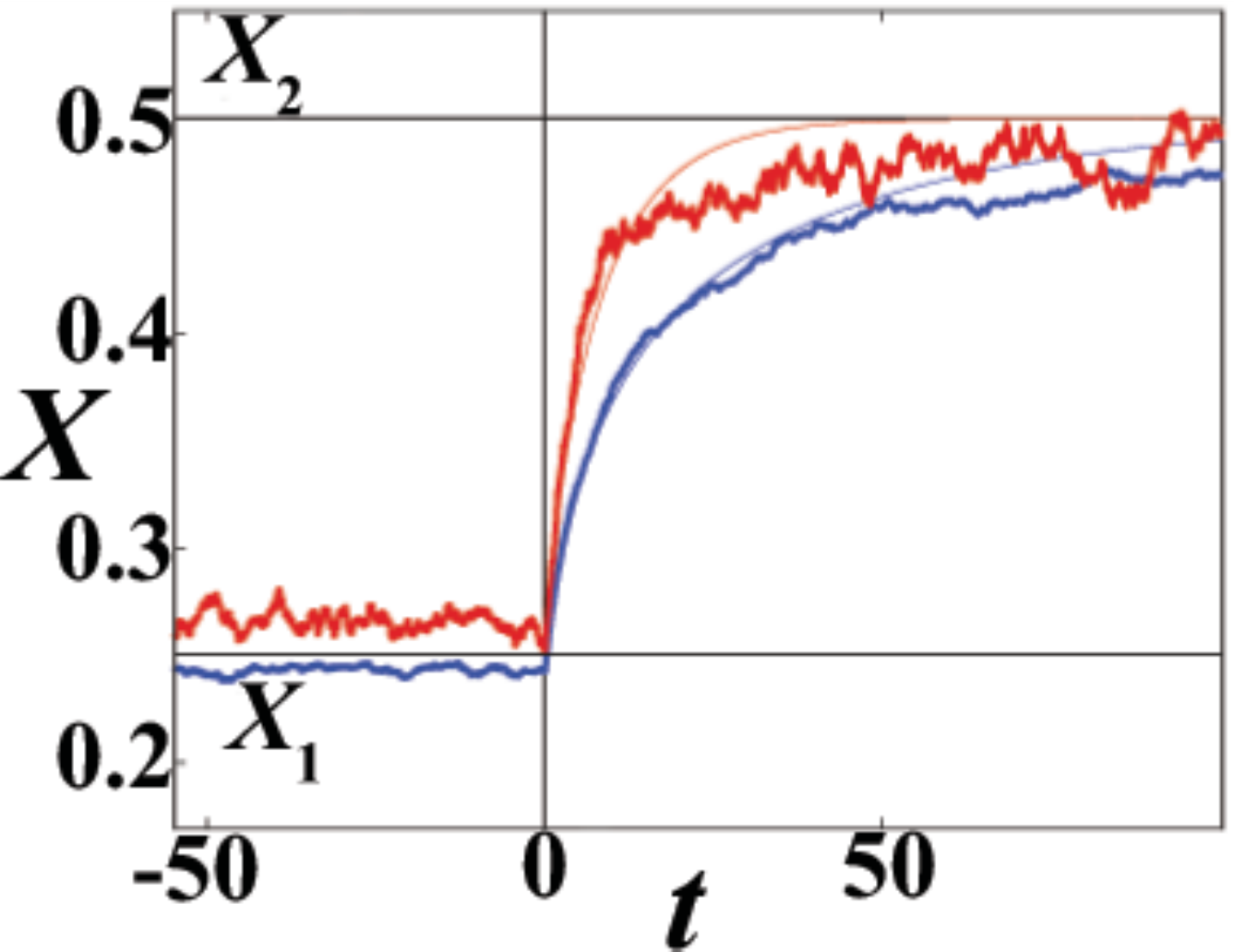}
\protect\protect\caption{(Color online) Dependence of the transient dynamics of the network
on $B$ . The change of external input occurs at $t=0$. The response
of the network for $B=0$ is shown by the blue (light gray) lines,
and for $B=0.01$ by the red (dark gray) lines. The remaining network
parameters are $N=400,$ $\alpha=0.65$, $D=0.001$. . The thick lines
represent the numerical results, whereas the thin lines denote the
theoretical estimates.\label{fig:trans}}
\end{figure}

Thus, strengthening of the external noise $B$ increases the rate
$\Gamma$ and speeds up the network response. This finding is corroborated
by numerical simulations illustrated in Fig. \ref{fig:trans} Here,
two networks are considered: the first one without the external noise
($B=0$, blue curve), and the second one with noise ($B=0.01$, red
curve). The values of the other parameters are given in the caption
to the figure. For both cases, the input $I$ changes its value at
the moment $t=0$ so that the stationary value of $X$ changes from
$X_{1}=0.25$ to $X_{2}=0.5$. The estimate (\ref{eq:rate_thermo})
then gives $\Gamma=0.025$ without noise and $\Gamma=0.103$ with
noise, which implies a fourfold speedup of the network response. Note
that the numerical results show satisfactory agreement with the theoretical
predictions. 

\section{Finite size effects\label{sec:Finite}}

In this section we analyze the influence of the finite-size effects
in case where the network is large but finite, viz. $N\gg1$. Then,
the noise term in \eqref{eq:R} can no longer be considered zero,
and can give rise to stochastic fluctuations of the mean rate around
the values obtained for the thermodynamical limit.

To study the magnitude of fluctuations, let us rewrite \eqref{eq:R}
as follows: 
\begin{equation}
\frac{dX}{dt}=F(X)+\frac{1}{N}G(X,S)+\frac{1}{\sqrt{N}}\sqrt{2\Psi(X,S)}\eta.\label{eq:X}
\end{equation}
For large $N$, the variables $X$ and $S$ are close to the respective
values $X_{0}$ and $S_{0}$ from the thermodynamic limit, whereby
$X_{0}$ is defined by the condition $F(X_{0})=0$, and $S_{0}$ by
(\ref{eq:S0}). Since the fluctuations $x=X-X_{0}$ are small, one
can linearize \eqref{eq:X} and obtain 
\begin{equation}
\frac{dx}{dt}=F^{\prime}(X_{0})x+\frac{1}{N}G(X_{0},S_{0})+\frac{1}{\sqrt{N}}\sqrt{2\Psi(X_{0},S_{0})}\eta.\label{eq:X_lin}
\end{equation}
Since the state is stable in the thermodynamic limit, $F^{\prime}(X_{0})<0$
applies. The steady state's displacement due to the finite size effect
equals 
\begin{equation}
x_{0}=\frac{G(X_{0},S_{0})}{-NF^{\prime}(X_{0})}=\frac{H_{2}\left(\mu(X_{0}-I)^{2}+\alpha^{2}p\left(D+BH_{1}^{2}\right)\right)}{Np^{2}\left(12\alpha B+1-\alpha H_{1}\right)}.\label{eq:x0}
\end{equation}
This deviation is of the order of $1/N$, while the random fluctuations
of $X$ due to noise are of the order of $1/\sqrt{N}$. This allows
one to neglect the second term in \eqref{eq:X_lin} and obtain the
following expression for the variance of $X$ over stochastic realizations:
\begin{equation}
\left[x{}^{2}\right]=\frac{2\Psi(X_{0},S_{0})}{-2NF^{\prime}(X_{0})}=\frac{\left(D+BH_{1}^{2}\right)\left(2+\alpha^{2}H_{1}^{2}\right)}{2N\left(12\alpha B+1-\alpha H_{1}\right)}.\label{eq:x2}
\end{equation}
The expressions (\ref{eq:x0}) and (\ref{eq:x2}) both contain $F'(X_{0})$
in the denominator. When the value of $F'(X_{0})$ becomes small,
the two formulas lose validity since the linearization of \eqref{eq:X})
is no longer adequate. Note that such scenario corresponds to the
parameter domain near the saddle-node bifurcations of the system in
the thermodynamic limit.

\begin{figure*}[t]
\centering \hspace{-0.3cm} \includegraphics[scale=0.165]{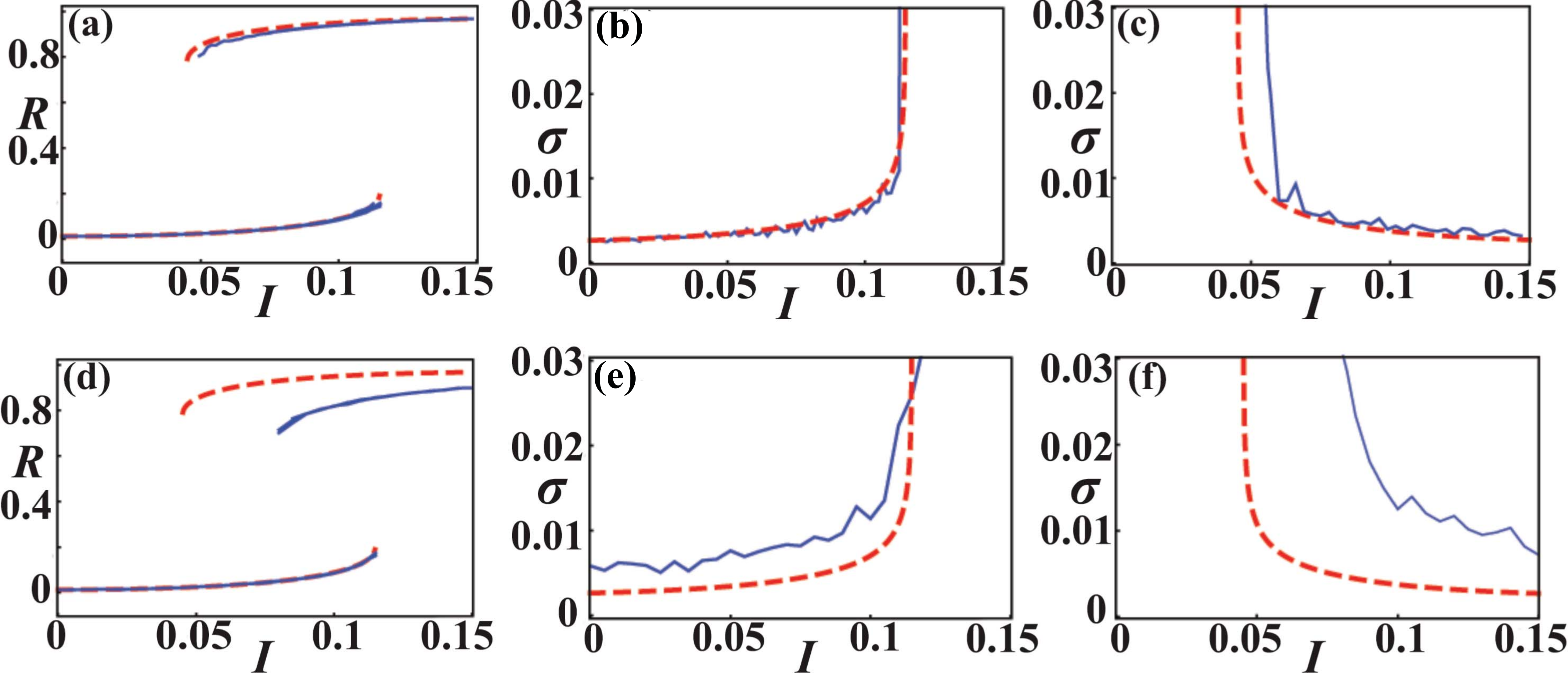}
\protect\protect\caption{(Color online)(a) One-parameter bifurcation diagram illustrating the
dependence of the stationary mean rate $R$ versus the bias current
$I$. Blue solid lines indicates the values observed after the transient,
while red dashed lines show the theoretical predictions for the stable
levels. (b) and (c) The variance of the mean rate $\sigma=\sqrt{[\delta X^{2}]}$.
The solid line denotes the numerical results, whereas the dashed line
stands for the approximate model. The parameter values are $N=400$,
$p=0.2$, $c=4.2$, $B=D=0.002$. The second row is intended to illustrate
the breakdown of theory at smaller network sizes. The presentation
style is the same as in the upper row, but the size of the exact system
is $N=70$. (d), (e) and (f) show that the stationary mean-rate, as
well as the associated variance substantially depart from what is
predicted by the approximate model. \label{fig:finite}}
\end{figure*}

In order to verify the validity and the accuracy of the developed
mean-field approach, we have performed direct simulations of the network
(\ref{eq:1}) and compared the results with the predictions of the
theory. We find that for $B<0.01$, $D<0.01$ and $N>100$ the theory
holds quite well in most of the cases: the mean rate of the network
is typically predicted with the accuracy no less than $5$\%. The
theory's validity reduces for the values of $R$ close to zero and
unity since the second derivative $H_{2}$ has discontinuity at these
points.

The comparison between the numerical and the theoretical results is
provided in Fig. \ref{fig:finite}. The intention is to first consider
a sufficiently large network $N=400$, where the mean-field treatment
is expected to hold, see Fig. \ref{fig:finite}(a). In particular,
for each parameter value, the network is simulated for the period
$T=200$ starting from $10$ different randomly chosen initial conditions.
After the transient $T_{tr}=50$, all the observed mean rates $R$
were saved and plotted versus the corresponding parameter value. The
theoretical prediction for the mean is superimposed on this plot (see
the dashed lines). To check the predictions for the magnitude of the
stochastic fluctuations, we have further plotted together the observed
variance and the estimate \eqref{eq:x2}, cf. Figs. \ref{fig:finite}(b),(c).
Since the network is bistable in a certain parameter interval, the
results are plotted separately for the low and the high branches.
As expected, the theory becomes inadequate close to the points where
the branches vanish through the saddle-node bifurcations. In the rest
of the parameter interval the theoretical estimate is quite precise.

The second row in Fig. \ref{fig:finite} illustrates the breakdown
of theory for smaller system sizes. As an example, we consider the
case $N=70$. Note that the upper branch of the mean rates substantially
deviates from the theoretical prediction. One also finds that the
magnitude of stochastic fluctuations are much larger than what is
anticipated by the approximate model, because the assumptions behind
\eqref{eq:x2} no longer hold.

Note that the influence of the system finite-size on the value of
the variance $S$ amounts only to its small change, which is of the
order $1/N$. Namely, the stationary value of the variance for large
$N$ equals 
\begin{equation}
S=S_{0}+\frac{p(1-p)c^{2}}{2N}\left(B(8H^{2}R^{2}-H_{1}^{2})-DH_{1}^{2}\right).
\end{equation}

\section{Summary and discussion\label{sec:Summary}}

In this paper, we have considered a network of rate-based neurons
with random connectivity and two types of noise. In order to study
the macroscopic dynamics of the network, we have developed the second-order
mean-field approach which incorporates the Gaussian closure hypothesis.
The dynamics of the large, but finite network is described in terms
of the assembly averaged firing rate and the associated variance,
whose evolution is given by the system \eqref{eq:R}-\eqref{eq:S}.
The main approximations relevant to the derivation of the model are
that the outputs of the units are unbiased and uncorrelated. The analysis
shows that these assumptions are valid for large networks with random
sparse connectivity.In fact, such type of connectivity renders correlation
between the outputs of units small, which is the point relevant for
our derivation.

In the context of neuroscience, random networks are often considered
as the simplest model of connectivity of neural circuits \cite{BH99,B00,TTF12,VS96,VS98}.
On the other hand, most of the research so far dedicated to mean-field
approach for stochastic systems has addressed the scenario of a fully
connected network \cite{T12,LGNS04,FTVB13,H07a,H07b,H09}. However,
recent experimental data provides evidence that the organization of
synaptic connections in brain is nontrivial and differs drastically
from both of the above models \cite{M97,SSRNC05,LTSP09,PBM11}. The
structure of neural networks appears to be inhomogeneous, in a sense
that most of the connections are random and sparse, but some units
are also organized into densely connected clusters \cite{KTNF14,LKD12}. 

Such clusters have already been established to play an important information-processing
role in the cortex \cite{ZZHK06,KCB11,YDC05,YBJ10}. Within a broader
research agenda, the results gained here for the case of random networks,
if incorporated together with the previous work on fully connected
networks, may ultimately allow us to derive the mean-field model appropriate
for clustered networks.

In terms of research goals, most of the early studies applying the
mean-field approach have been focused on explaining the mechanisms
behind the spontaneous activity characterized by irregular firing
of neurons at low rates, typically found in a living cortex or living
hippocampus. Apart from gaining insight into the genesis and the self-sustaining
property of these chaotic states, the aim has also been to explain
why populations of highly nonlinear units display linear responses
to external drive, reacting on time scales faster than the characteristic
time scale of a single unit. The emergence of relevant cooperative
states has been linked to several different ingredients, including
the features of the unit\textquoteright s threshold function, the
network connection topology and the scaling of synaptic strengths.
In particular, for a fully connected network of rate-based units with
random asymmetrical couplings similar to spin glasses, the onset of
chaos has been associated to the gain parameter of the threshold function
\cite{SCS88}. For networks comprised of binary neuron-like units,
the most important finding has concerned the existence of a chaotic
balanced state, where variability is achieved by the balance of excitatory
and inhibitory inputs, each being much larger than the unit\textquoteright s
threshold \cite{VS96,VS98}. Necessary conditions for maintaining
such a regime include random and sparse connectivity, as well as comparably
strong synapses. Under similar conditions the networks of integrate-and-fire
neurons have been found to support a bistable regime between the spontaneous
activity, uncorrelated with the received stimuli, and the \textquotedblleft working
memory\textquotedblright{} states, strongly correlated with the \textquotedblleft learned\textquotedblright{}
stimuli \cite{AB97a,AB97b}. Further research have revealed importance
of the weight distribution in random networks of integrate-and-fire
neurons and its essential role for the spike-based communication \cite{TTF12}.

At variance with theabove models, which typically do not consider
at all or provide only a limited account of the effects of noise,
the central issue of research in recent years has become the point
of how noise from the level of single units is translated to and reflected
in the macroscopic-scale behavior. The present study aims to contribute
to this line of research, and our main results can be summarized as
follows. In the thermodynamic limit, the network dynamics is deterministic
in nature. We have determined the stationary levels of the network
activity, showing that for strong enough coupling ($\alpha>\alpha_{0}$,
see Eq. (\ref{eq:pitchfork})) the network exhibits bistable regime,
characterized by coexistence of the low and the high stable states.
In terms of how noise from microscopic dynamics effectively impacts
the collective behavior, our most important finding is that the external
and the internal noise play essentially different roles in the mean-field
dynamics. In particular, in the thermodynamic limit, the internal
noise does not influence the macroscopic dynamics at all, while the
external noise changes the position and the number of stable levels.
We have demonstrated that this feature can be used to control the
network dynamics via external noise in a hysteresis-like scenario,
as illustrated in Fig. \ref{fig:noise}. We have also shown that the
external noise influences the transient dynamics of a network,, at
certain instances being able to speed up its response to the change
of external drive..

The developed theory has also allowed us to consider the finite-size
effects on the network dynamics. The corresponding approximate model
for large but finite networks effectively involves three sources of
noisy behavior. Apart from the internal and external noises, which
manifest as the additive and multiplicative noise at the macroscopic
level, we identify an additional term that derives from heterogeneity
in the units' connectivity degrees. We have found that the finite-size
effects are twofold and consist in $(i)$ displacement of the stationary
levels and $(ii)$ in giving rise to stochastic fluctuations of the
mean rate. Since the change of the stationary values of $R$ and $S$
is of the order of $1/N$, the most important are the stochastic fluctuations
which have the magnitude of the order of $1/\sqrt{N}$. It has also
been explicitly demonstrated that the developed approach provides
satisfactory estimate of the magnitude of the fluctuations for the
parameter domain sufficiently away from the bifurcations.

We suspect that novel interesting effects may arise in sufficient
vicinity of the pitchfork bifurcation, where the network possesses
two stable activity levels that are relatively close to each other.
In this case, the derivatives $F^{\prime}(X_{0})$ are close to zero
for both states, and the estimate provided by Eq.\eqref{eq:x2} indicates
large fluctuations of the mean rate. If one approaches close enough
to the bifurcation, the magnitude of fluctuations may become of the
order of the distance between the levels, which is likely to induce
stochastic ``switching'' between the low and the high state. This
phenomenon may be associated to high variability of firing rates often
observed in neural networks and recently connected to clustering of
synaptic connections \cite{LKD12}. However, linearization of Eq.
\eqref{eq:X} in this case is no longer adequate, such that the full
nonlinear equations (\ref{eq:R}-\ref{eq:S}) should be studied to
capture the potential phenomenon of stochastic switchings. This will
be one of the main goals for our future research.

\section*{Acknowledgments}

This work was supported in part by the Russian Foundation for Basic
Research (grant 14-02-00042) and the Ministry of Education and Science
of the Republic of Serbia under project No. $171017$. The authors
would like to thank Prof. Nikola Buri\'{c} and Prof. Vladimir Nekorkin
for valuable discussions during the different stages of the research.


\begin{thebibliography}{10}
\bibitem{SK93}W. R. Softky, C. Koch, J. Neurosci. \textbf{13}, 334
(1993).

\bibitem{STG01}M. Steriade, I. Timofeev, F. Grenier, J. Neurophysiol.
\textbf{85}, 1969 (2001).

\bibitem{DRP03}A. Destexhe, M. Rudolph, D. Par�, Nat. Rev. Neurosci.
\textbf{4}, 739 (2003).

\bibitem{B06a}G. Buzs�ki, \emph{Rhythms of the Brain}, (Oxford University
Press, New York, 2006).

\bibitem{M57}V. B. Mountcastle, Brain \textbf{120} 701 (1957).

\bibitem{GZTK08} P. beim Graben, C. Zhou, M. Thiel, and J. Kurths,
eds., \emph{Lectures in Supercomputational Neuroscience: Dynamics
in Complex Brain Networks}, (Springer-Verlag, Berlin Heidelberg, 2008).

\bibitem{CGPW14} S. Coombes, P. beim Graben, R. Potthast and J. Wright,
eds., \emph{Neural Fields, Theory and Applications}, (Springer-Verlag,
Berlin Heidelberg, 2014).

\bibitem{WC72}H. R. Wilson, and J. D. Cowan, Biophys. J. \textbf{12},
1 (1972).

\bibitem{WC73}H. R. Wilson, and J. D. Cowan, Kybernetik \textbf{13},
55 (1973).

\bibitem{A77}S. Amari, Biol. Cybern. \textbf{27}, 77 (1977).

\bibitem{RRWBGR01}P. A. Robinson, C. J. Rennie, J. J. Wright, H.
Bahramali, E. Gordon, and D. L. Rowe, Phys. Rev. E \textbf{63} 021903
(2001).

\bibitem{JH97}V. K. Jirsa, and H. Haken, Phys. D \textbf{99} 503
(1997).

\bibitem{LC01}C. R. Laing, and C. C. Chow, Neural Comput. \textbf{13},
1473 (2001).

\bibitem{LTGE02}C. R. Laing, W. C. Troy, B. Gutkin, and G. B. Ermentrout,
SIAM J. Appl. Math. \textbf{63}, 62 (2002).

\bibitem{PE01}D. J. Pinto and G. B. Ermentrout, SIAM J. Appl. Math.
\textbf{62}, 206 (2001).

\bibitem{FB05b}S. E. Folias and P. C. Bressloff, SIAM J. Appl. Math.
\textbf{65} 2067 (2005).

\bibitem{EB03}M. Enculescu and M. Bestehorn, Phys. Rev. E \textbf{67},
041904 (2003).

\bibitem{L05}C. R. Laing, SIAM J. Appl. Dyn. Syst. \textbf{4}, 588
(2005).

\bibitem{FB05a}S. E. Folias, and P. C. Bressloff, Phys. Rev. Lett.
\textbf{95}, 208107 (2005).

\bibitem{OLC07}M. R. Owen, C. R. Laing, and S. Coombes, New J. Phys.
\textbf{9}, 378 (2007).

\bibitem{B09}P. C. Bressloff, SIAM J. Appl. Math. \textbf{70}, 1488
(2009).

\bibitem{B10}P. C. Bressloff, Phys. Rev. E \textbf{82}, 051903 (2010).

\bibitem{T12}J. Touboul, Physica D \textbf{241} 1223 (2012).

\bibitem{TE11}J. D. Touboul, and G. B. Ermentrout, J. Comput. Neurosci.
\textbf{31}, 453 (2011).

\bibitem{RMBWP07}A. Renart, R. Moreno-Bote, X. J. Wang, and N. Parga,
Neural Comput. \textbf{19}, 1 (2007).

\bibitem{MBP06}R. Moreno-Bote, and N. Parga, Phys. Rev. Lett. \textbf{96},
028101 (2006).

\bibitem{RD14}R. Rosenbaum, and B. Doiron, Phys. Rev. X \textbf{4},
021039 (2014).

\bibitem{FSW08}A. A. Faisal, L. P. J. Selen, and D. M. Wolpert, Nat.
Rev. Neurosci. \textbf{9}, 292 (2008).

\bibitem{DW11}M. D. McDonnell, and L. M. Ward, Nat. Rev. Neurosci.
\textbf{12}, 415 (2011).

\bibitem{AANVSG07}V. S. Anishchenko, V. Astakhov, A. Neiman, T. Vadivasova,
L. Schimansky-Geier, \emph{Nonlinear Dynamics of Chaotic and Stochastic
Systems: Tutorial and Modern Developments}, (Springer-Verlag, Berlin
Heidelberg, 2007).

\bibitem{HJBS06}B. Hauschildt, N. B. Janson, A. Balanov, and E. Sch�ll,
Phys. Rev. E \textbf{74}, 051906 (2006).

\bibitem{JBS08}N. B. Janson, A. G. Balanov, and E. Sch�ll, ``Control
of Noise-Induced Dynamics'' in \emph{Handbook of Chaos Control},
(eds. E. Sch�ll, and H. G. Schuster, Wiley, 2008, 2nd ed.), p. 221-273.

\bibitem{LGNS04}B. Lindner, J. Garcia-Ojalvo, A. Neiman, and L. Schimansky-Geier,
Phys. Rep. \textbf{392}, 321 (2004).

\bibitem{FTVB13}I. Franovi\'{c}, K. Todorovi\'{c}, N. Vasovi\'{c},
and N. Buri\'{c}, Phys. Rev. E \textbf{87}, 012922 (2013).

\bibitem{FTVB14}I. Franovi\'{c}, K. Todorovi\'{c}, N. Vasovi\'{c},
and N. Buri\'{c}, Phys. Rev. E \textbf{89}, 022926 (2014).

\bibitem{HLSG08}A. Hutt, A. Longtin, and L. Schimansky-Geier, Physica
D \textbf{237}, 755 (2008).

\bibitem{FTC09}O. Faugeras, J. Touboul, and B. Cessac, Front. Comput.
Neurosci. \textbf{3}, 1 (2009).

\bibitem{BD09}S. E. Boustani, and A. Destexhe, Neural Comput. 21,
46100 (2009).

\bibitem{K92}N. G. Van Kampen, \emph{Stochastic Processes in Physics
and Chemistry}, (North-Holland, Amsterdam, 1992).

\bibitem{BC07}M. A. Buice, and J. D. Cowan, Phys. Rev. E \textbf{75},
051919 (2007).

\bibitem{BCC10}M. Buice, J. D. Cowan, and C. C. Chow, Neural Comput.
\textbf{22}, 377 (2010).

\bibitem{GH93}W. Gerstner, J. L. van Hemmen, Phys. Rev. Lett. \textbf{71},
312 (1993).

\bibitem{BH99}N. Brunel, and V. Hakim, Neural Comput. \textbf{11},
1621 (1999).

\bibitem{B00}N. Brunel, J. Comput. Neurosci. \textbf{8}, 183 (2000).

\bibitem{AV93}L. F. Abbott, C. van Vreeswijk, Phys. Rev. E \textbf{48},
1483 (1993).

\bibitem{NT00}D. Nykamp, D. Tranchina, J. Comput. Neurosci. \textbf{8},
19 (2000).

\bibitem{LT07}C. Ly, D. Tranchina, Neural Comput. \textbf{19}, 2032
(2007).

\bibitem{SC07}H. Soula and C. C. Chow, Neural Comput. \textbf{19},
3262 (2007).

\bibitem{GS94}I. Ginzburg, and H. Sompolinsky, Phys. Rev. E \textbf{50},
3171 (1994).

\bibitem{MV02}C. Meyer, and C. van Vreeswijk, Neural Comput. \textbf{14},
369 (2002).

\bibitem{H07a}H. Hasegawa, Phys. Rev. E \textbf{75}, 051904 (2007).

\bibitem{H07b}H. Hasegawa, ``A Generalized Rate Model for Neuronal
Ensembles'' in \emph{Neuronal Network Research Horizons}, (ed. M.
L. Weiss, Nova Science Publishers, 2007), p. 61-98.

\bibitem{H09}H. Hasegawa, Physica A \textbf{388}, 499 (2009).

\bibitem{UR99}W. M. Ursey, and R. C. Reid, Annu. Rev. Physiol. \textbf{61},
435 (1999).

\bibitem{CZ00}R. C. deCharms and A. Zador, Annu. Rev. Neurosci. \textbf{23},
613 (2000).

\bibitem{AMP04}R. A. Anderson, S. Musallam, and B. Pesaran, Curr.
Opin. Neurobiol. \textbf{14}, 720 (2004).

\bibitem{CMMN99}J. K. Chapin, K. A. Moxon, R. S. Markowitz, and M.
A. L. Nicolelis, Nat. Neurosci. \textbf{2}, 664 (1999).

\bibitem{H03}H. Hasegawa, Phys. Rev. E \textbf{67}, 041903 (2003).

\bibitem{FB02}N. Fourcaud, and N. Brunel, Neural Comput. \textbf{14},
2057 (2002).

\bibitem{SHS03}O. Shriki, D. Hansel, and H. Sompolinsky, Neural Comput.
\textbf{15}, 1809 (2003).

\bibitem{M97}H. Markram, Cereb. Cortex \textbf{7}, 523 (1997).

\bibitem{SSRNC05} S. Song, P. J. Sj�str�m, M. Reigl, S. Nelson, and
D. B. Chklovskii, PLoS Biol. \textbf{3}, e68 (2005).

\bibitem{ZZHK06}C. Zhou, L. Zemanova, G. Zamora, C. C. Hilgetag,
and J. Kurths, Phys. Rev. Lett. \textbf{97}, 238103 (2006).

\bibitem{LTSP09}S. Lefort, C. Tomm, J.-C. F. Sarria, C. C. H. Petersen,
Neuron \textbf{61}, 301 (2009).

\bibitem{PBM11}R. Perin, T. K. Berger, and H. Markram, Proc. Natl.
Acad. Sci. \textbf{108}, 5419 (2011).

\bibitem{KTNF14}V. V. Klinshov, J. Teramae, V. I. Nekorkin, and T.
Fukai, PLoS One \textbf{9}, e94292 (2014).

\bibitem{TTF12}J. Teramae, Y. Tsubo, and T. Fukai, Sci. Rep. 2, (2012).

\bibitem{VS96}C. van Vreeswijk, and H. Sompolinsky, Science \textbf{274},
1724 (1996).

\bibitem{VS98}C. van Vreeswijk, and H. Sompolinsky, Neural Comput.
\textbf{10}, 1321 (1998).

\bibitem{T98}H. C. Tuckwell, \emph{Introduction to theoretical neurobiology},
(Cambridge University Press, 1998).

\bibitem{B06}A. N. Burkitt, Biol. Cybern. \textbf{95}, 1 (2006).

\bibitem{LKD12}A. Litwin-Kumar, and B. Doiron, Nat. Neurosci. \textbf{15},
1498 (2012).

\bibitem{KCB11}H. Ko, L. Cossell, C. Baragli, J. Antolik, C. Clopath,
S. B. Hofer and T. D. Mrsic-Flogel, Nature \textbf{473}, 87 (2011).

\bibitem{YDC05}Y. Yoshimura, J. L. M. Dantzker, and E. M. Callaway,
Nature \textbf{433}, 868 (2005).

\bibitem{YBJ10}L. Yassin L, B. L. Benedetti, J. S. Jouhanneau, J.
A. Wen, J. F. Poulet, and A. L. Barth, Neuron \textbf{68}, 1043 (2010).

\bibitem{SCS88} H. Sompolinsky, A. Crisanti, and H. J. Sommers, Phys.
Rev. Lett. \textbf{61}, 259 (1988). 

\bibitem{AB97a}D. J. Amit and N. Brunel, Netw. Comput. Neural Syst.
\textbf{8}, 373 (1997). 

\bibitem{AB97b} D. J. Amit and N. Brunel, Cereb. Cortex \textbf{7},
237 (1997). \end{thebibliography}
\end{document}